PAPER • OPEN ACCESS

# Parity violation in primordial tensor non-Gaussianities from matter bounce cosmology

To cite this article: Shingo Akama and Mian Zhu JCAP07(2024)039

View the article online for updates and enhancements.

## You may also like

- Trans-Planckian issues for inflationary cosmology
  Robert H Brandenberger and Jérôme Martin

- Finite-time correlations boost large voltage angle fluctuations in electric power grids
  Melvyn Tyloo, Jason Hindes and Philippe Jacquod

- THE HALO MASS FUNCTION FROM EXCURSION SET THEORY. I. GAUSSIAN FLUCTUATIONS WITH NON-MARKOVIAN DEPENDENCE ON THE SMOOTHING SCALE
  Michele Maggiore and Antonio Riotto

# Parity violation in primordial tensor non-Gaussianities from matter bounce cosmology

Shingo Akama® and Mian Zhu®

*Faculty of Physics, Astronomy and Applied Computer Science, Jagiellonian University, 30-348 Krakow, Poland*

*E-mail:* shingo.akama@uj.edu.pl, mian.zhu@uj.edu.pl

ABSTRACT: It has been shown that primordial tensor non-Gaussianities from a cubic Weyl action with a non-dynamical coupling are suppressed by the so-called slow-roll parameter in a conventional framework of slow-roll inflation. In this paper, we consider matter bounce cosmology in which the background spacetime is no longer quasi-de Sitter, and hence one might expect that the matter bounce models could predict non-suppressed non-Gaussianities. Nevertheless, we first show that the corresponding non-Gaussian amplitudes from the cubic Weyl term with a non-dynamical coupling are much smaller than those from the conventional slow-roll inflation, in spite of the fact that there is no slow-roll suppression. We then introduce a dynamical coupling that can boost the magnitude of graviton cubic interactions and clarify that there is a parameter region where the tensor non-Gaussianities can be enhanced and can potentially be tested by cosmic microwave background experiments.

https://doi.org/10.1088/1475-7516/2024/07/039

# Contents

## 1 Introduction

Recent observations suggest the existence of parity violation physics. For instance, a nearly frequency-independent cosmic birefringence (i.e., the rotation of the plane of cosmic microwave background (CMB) linear polarization) has been measured by the Planck data [1, 2]. A joint analysis from Planck and WMAP data implies a non-zero polarization angle $\beta = 0.34 \pm 0.09°$ [3]. A tentative evidence for parity-violating scalar trispectrum on large scale structure (LSS) is reported by using the BOSS CMASS galaxy sample [4, 5]. More precise evidence for parity violation will be available from upcoming astrophysical and cosmological experiments. It is thus interesting to explore theoretical possibilities causing parity violation. Since the CMB and LSS are believed to have their origin in primordial fluctuations, it is natural to ask if the parity violation results from the early universe (see [5–7] for relevant discussions).

Inflation [8–10] is the leading paradigm of the early universe, and parity violation in this paradigm has been widely studied [11–41]. Recent studies have shown the absence of parity violation in tree-level cosmological correlation functions of scalar perturbations for vanilla single-field inflation [27, 38]. In addition, a standard slow-roll inflation typically predicts suppressed parity-violating signatures in well-known parity-violating gravitational theories [42–44]. Apart from the observational aspects, inflation is plagued with the conceptional problems such as the initial singularity problem [45–47] and the Trans-Planckian problem [48–51]. See ref. [52] for an intuitive explanation on how a classical non-singular bounce resolve fundamental problems in cosmology and also [53] for a critical review on problems of such scenarios. Also, by studying alternatives to inflation, we can observationally compare inflation with alternatives and test those paradigms, especially in terms of parity-violating signatures in the current context. The correct paradigm of the early universe will be clarified by testing and distinguishing possible scenarios, and hence at this stage it is still worth exploring alternative scenarios. We are thus motivated to explore observable parity-violating signatures in non-singular alternatives [54, 55] in which the conceptional problems are resolved.

From the viewpoint of an effective field theory (EFT) approach, a parity-violating higher-curvature correction to the Einstein-Hilbert term at the leading order is a topological term called the Chern-Simons term, which is of $\mathcal{O}(\mathcal{R}^4)$ with $\mathcal{R}$ being the spacetime curvature. A non-topological term is obtained by introducing a dynamical coupling at the cost of a ghost degree of freedom [56, 57].[1] Furthermore, parity-violating signatures in the cosmological correlation functions from the dynamical Chern-Simons gravity are generically suppressed in proportion to $H/M_{\rm CS}$ where $H$ is the Hubble parameter and $M_{\rm CS}$ is the scale at which the ghost appears [43, 44]. We thus work on the next-to-leading-order correction (i.e., the cubic Weyl action of $\mathcal{O}(\mathcal{R}^6)$) as the simplest ghost-free parity-violating gravitational theory.

Since the cubic Weyl term is nonvanishing on cosmological backgrounds from the cubic order in perturbations, the parity-violating signatures start to appear at the bispectrum level. In the cubic Weyl action, the primordial non-Gaussianities have been well investigated in the context of inflation in the literature. In particular, it has been shown in ref. [42] that there is no parity violation in a three-point correlation function of graviton when the following three conditions hold: (1) the background spacetime is exact-de Sitter, (2) the coupling function of the cubic Weyl term is time independent, and (3) the initial vacuum state of graviton is the Bunch-Davies one. Nonvanishing parity violation has been obtained by breaking the condition (1) in ref. [42], the condition (2) in ref. [59], and the condition (3) in ref. [39]. Also, the authors of ref. [42] have computed the three-point function of graviton from slow-roll inflation with the Bunch-Davies initial state and found that the resultant magnitude is proportional to the slow-roll parameter $\epsilon \equiv -H^{-2}dH/dt$ and is thus suppressed for $\epsilon \ll 1$ in the quasi-de Sitter inflation.

In the present paper, we explore the possibility to obtain enhanced tensor non-Gaussianities at the bispectrum level from the non-singular alternatives to inflation in the context of cubic Weyl action. Specifically, we will work in matter bounce cosmology [60] in which a scale-invariant scalar power spectrum is obtained from a matter-dominated contracting phase. We first show that the primordial tensor non-Gaussianities generated during the matter-dominated contracting phase cannot be enhanced even if the condition (1) is absent and the parameter $\epsilon = 3/2$ is larger than that in inflation. This is because cubic interaction terms of the tensor perturbations are suppressed after horizon crossing, after which the primordial non-Gaussianities are generated. In light of this, we introduce a dynamical coupling to cancel that suppression, which breaks the condition (2). As a result, we clarify that the non-Gaussianities can be enhanced for certain cases, which can potentially be tested, especially by upcoming CMB experiments.

This paper is organized as follows. In section 2, we introduce the setup for our work, including the matter bounce cosmology and the cubic Weyl action. In section 3, we present the result of the three-point correlation function for the non-dynamical and dynamical cases. In the same section, we investigate the parameter region for the enhanced tensor bispectrum without strong coupling problems. Our conclusion is drawn in section 4.

---

[1]This is true unless one regards it as a low-energy effective field theory of some fundamental theory or adopts the so-called unitary gauge, i.e., a scalar field is homogeneous on a time-constant hypersurface [57]. See also ref. [58] for healthy extensions without the instability.

## 2 Setup

We work in a spatially-flat Friedmann-Lemaître-Robertson-Walker (FLRW) spacetime whose metric is of the form,

$$ds^2 = -dt^2 + a(t)^2 dx_i dx^i = a(\tau)^2(-d\tau^2 + dx_i dx^i), \tag{2.1}$$

where $a$ is the scale factor and $\tau$ denotes the conformal time defined by $d\tau = dt/a$. In the rest of this paper, we use a dot and prime to denote differentiation with respect to the cosmic time $t$ and the conformal time $\tau$, respectively.

Matter bounce cosmology, where a matter-dominated contracting phase is followed by a bouncing phase and a subsequent expanding one, can predict a nearly scale-invariant primordial power spectra, consistent with the CMB experiments [61] (see e.g., [62–65] for concrete realizations of a matter contracting phase). For simplicity, we take the matter contraction phase to be described by a scalar field minimally coupled to general relativity (that is naturally parity-preserving):

$$S = \int d\tau d^3x \sqrt{-g}(\mathcal{L}_{\mathrm{EH}} + \mathcal{L}_\phi), \tag{2.2}$$

where $\mathcal{L}_{\mathrm{EH}}$ is the Einstein-Hilbert term and $\mathcal{L}_\phi$ is the Lagrangian density of the scalar field. It has been shown in ref. [66] that a matter bounce scenario consistent with observations can be realized from the action (2.2), as long as $\mathcal{L}_\phi$ contains the cubic Galileon term $G(\phi, -(\partial_\mu \phi)^2/2) \Box \phi$.

The evolution of the statistical properties of the perturbations (even whether those change or not) during the subsequent bouncing and expanding phases is model dependent [67–70]. For simplicity, we shall restrict ourselves to the contracting phase by assuming that the impacts from the subsequent phases are negligible to the tensor non-Gaussianities generated during the contracting phase.[2] Also, in general, the non-singular cosmological solutions are plagued with gradient instabilities in scalar perturbations if the entire history of cosmic evolution is described by the Horndeski theory [71–73]. In the present paper, we just assume that at least the contracting phase is described by the cubic-Galileon theory and some beyond Horndeski operator [74–80] is brought into play somewhere away from the contracting phase.

The scalar factor during the matter-dominated universe scales as $a \propto \tau^2$, and let us parametrize the scale factor during the contracting phase as

$$a(\tau) = \left(\frac{-\tau}{-\tau_0}\right)^2, \quad -\infty < \tau < \tau_0 < 0, \tag{2.3}$$

where $\tau_0$ is the conformal time at the end of the contracting phase, and we normalized the scale factor by $a(\tau_0) = 1$. The Hubble parameter $H := \dot{a}/a$ evolves in time as

$$H = \frac{2}{3t} = \frac{2\tau_0^2}{\tau^3}. \tag{2.4}$$

---

[2]This assumption is indeed non-trivial. The non-Gaussianities generated during the bouncing phase would be subleading compared to those generated during the contracting phase as long as the time scale of bouncing phase is very short (e.g., such that the time evolution of the perturbations are negligible). Also, one can have non-trivial non-Gaussianities from the bouncing phase if the duration of bouncing phase is long enough (see the argument in ref. [68]). Thus, our assumption is valid for specific bouncing scenarios, e.g., bouncing models with a short bounce followed by an instantaneous reheating and then the radiation dominated epoch in the standard cosmic history.

As a parity-violating part, we consider the following cubic Weyl action

$$S_{\rm pv} = \int {\rm d}\tau d^3x \sqrt{-g} f(\phi) \epsilon^{\mu\nu\lambda\rho} W_{\mu\nu}^{\alpha\beta} W_{\alpha\beta}^{\gamma\sigma} W_{\lambda\rho\gamma\sigma}, \qquad (2.5)$$

where the Weyl tensor is defined by

$$W_{\mu\nu\rho\sigma} = R_{\mu\nu\rho\sigma} + \frac{1}{2}\left(R_{\mu\sigma}g_{\nu\rho} - R_{\mu\rho}g_{\nu\sigma} + R_{\nu\rho}g_{\mu\sigma} - R_{\nu\sigma}g_{\mu\rho}\right) + \frac{R}{6}\left(g_{\mu\rho}g_{\nu\sigma} - g_{\mu\sigma}g_{\nu\rho}\right), \quad (2.6)$$

and $\epsilon^{\mu\nu\lambda\rho}$ is the four-dimensional Levi-Civita symbol. Since the Weyl tensor vanishes at the background level, the cubic Weyl term does not affect the dynamics of either the contracting background or linear perturbations. For simplicity, we parametrize $f(\phi)$ as

$$f(\phi) = \frac{b}{M_p^2}\left(\frac{\tau}{\tau_0}\right)^\lambda, \qquad (2.7)$$

where $b$ is a dimensionless constant. Once one constructs a background solution, one can obtain the coupling function of the above form. For instance, in the cubic Galileon theory, a power-law model has been constructed with the scalar field satisfying $e^{\lambda\phi} \propto 1/(-t)$ where $\lambda$ is a constant [66]. Hence, the power-law coupling corresponds to a power of $e^{\lambda\phi}$. As another example, the scalar field with a power-law time dependence has been used in ref. [81]. In this case, the power-law coupling just corresponds to a power of $\phi$.

The tensor perturbations $h_{ij}$ are defined as

$$ds^2 = a^2(\tau)[-d\tau^2 + (\delta_{ij} + h_{ij})dx^i dx^j], \qquad (2.8)$$

where $h_{ij}$ obeys the transverse-traceless conditions, i.e., $h_{ii} = 0$ and $\partial^j h_{ij} = 0$. By expanding (2.2) up to quadratic order in $h_{ij}$, the quadratic action reads

$$S_h^{(2)} = \frac{M_p^2}{8}\int d\tau d^3x\, a^2 \left[h_{ij}'^2 - (\partial h_{ij})^2\right]. \qquad (2.9)$$

Note that the speed of gravitational waves is unity since $\phi$ is minimally coupled to gravity.

The tensor perturbations are quantized as

$$\hat{h}_{ij}(\vec{x},\tau) = \int \frac{d^3k}{(2\pi)^{3/2}} \hat{h}_{ij}(\vec{k},\tau) e^{i\vec{k}\cdot\vec{x}}$$

$$= \int \frac{d^3k}{(2\pi)^{3/2}} \sum_s \left[h_{\vec{k}}(\tau) e_{ij}^{(s)}(\vec{k}) \hat{a}_s(\vec{k}) + h_{-\vec{k}}^*(\tau) e_{ij}^{(s)*}(-\vec{k}) \hat{a}_s^\dagger(-\vec{k})\right] e^{i\vec{k}\cdot\vec{x}}, \qquad (2.10)$$

where the creation and annihilation operators are normalized as

$$\left[\hat{a}_s(\vec{k}), \hat{a}_{s'}^\dagger(\vec{k}')\right] = \delta_{ss'}\delta(\vec{k} - \vec{k}'). \qquad (2.11)$$

We adopt circular polarizations for the polarizations tensor whose explicit form is given in appendix A. The properties of the polarization tensor are as follows,

$$e_{ii}^{(s)}(\vec{k}) = 0, \qquad k_j e_{ij}^{(s)}(\vec{k}) = 0, \qquad e_{ij}^{(s)}(\vec{k}) e_{ij}^{*(s')}(\vec{k}) = \delta_{ss'},$$
$$e_{ij}^{*(s)}(\vec{k}) = e_{ij}^{(s)}(-\vec{k}), \qquad \epsilon_{ijl}\frac{\partial}{\partial x_l}\left[e_{mj}^{(s)}(\vec{k}) e^{i\vec{k}\cdot\vec{x}}\right] = s k e_{im}^s(\vec{k}) e^{i\vec{k}\cdot\vec{x}}, \qquad (2.12)$$

with the subscript $s = \pm 1$ representing the two helicity states of graviton (i.e., the right- and left-handed circular polarizations).

The dynamical equation of the tensor perturbation in Fourier domain is

$$h_{\vec{k}}'' + 2\frac{a'}{a}h_{\vec{k}}' + k^2 h_{\vec{k}} = 0. \tag{2.13}$$

We have the following solution,

$$h_{\vec{k}}(\tau) = -\frac{\sqrt{2}i}{M_p}\frac{\tau_0^2}{\tau^3}k^{-3/2}(1 + ik\tau)e^{-ik\tau}, \tag{2.14}$$

where we imposed an adiabatic (Minkowski) vacuum initial condition to a canonically normalized tensor perturbation, $v_{\vec{k}} := aM_p h_{\vec{k}}/2$, as

$$\lim_{\tau \to -\infty} v_{\vec{k}} = \frac{e^{-ik\tau}}{\sqrt{2k}}. \tag{2.15}$$

Notably, the amplitudes of the tensor perturbations grow in proportion to $1/\tau^3$ on the superhorizon scale, $|k\tau| \ll 1$. This is in contrast to the quasi-de Sitter inflation case where those are frozen on the superhorizon scales. The tensor power spectrum is defined by

$$\langle 0|\hat{h}^{(s)}(\vec{k})\hat{h}^{(s')*}(\vec{k}')|0\rangle = (2\pi)^3\delta^{(3)}(\vec{k} + \vec{k}')\delta_{ss'}\frac{\pi^2}{k^3}\mathcal{P}_h, \tag{2.16}$$

where $\hat{h}^{(s)}(\tau, \vec{k}) := \hat{h}_{ij}(\tau, \vec{k})e_{ij}^{(s)*}(\vec{k})$. The power spectrum evaluated at the end of the contracting phase at which the perturbations are on the superhorizon scales is then

$$\mathcal{P}_h = 2\frac{k^3}{2\pi^2}|h_k|^2 \simeq \frac{2}{\pi^2\tau_0^2 M_p^2} = \frac{H_0^2}{2\pi^2 M_p^2}, \tag{2.17}$$

where we denoted the Hubble parameter at $\tau_0$ as $H_0$, i.e., $H_0 = 2/\tau_0$.

## 3 Primordial tensor bispectra

We compute the three-point correlation function of $\hat{h}^{(s)}(\tau, \vec{k}) = \hat{h}_{ij}^{(s)}(\tau, \vec{k})e_{ij}^{(s)*}(\vec{k})$ as,

$$\langle \hat{h}^{(s_1)}(\tau, \vec{k}_1)\hat{h}^{(s_2)}(\tau, \vec{k}_2)\hat{h}^{(s_3)}(\tau, \vec{k}_3)\rangle = e_{i_1 j_1}^{(s_1)*}(\vec{k}_1)e_{i_2 j_2}^{(s_2)*}(\vec{k}_2)e_{i_3 j_3}^{(s_3)*}(\vec{k}_3)\langle \hat{h}_{i_1 j_1}^{(s_1)}(\vec{k}_1)\hat{h}_{i_2 j_2}^{(s_2)}(\vec{k}_2)\hat{h}_{i_3 j_3}^{(s_3)}(\vec{k}_3)\rangle. \tag{3.1}$$

Hereafter, we evaluate this quantity at the end of the contracting phase. The parity-preserving part of the three-point correlation function has been calculated in the context of matter bounce cosmology in refs. [66, 82], and hence in the present paper, we focus on the parity-violating part.

By employing the in-in formalism, one can compute the three-point function as

$$\langle \hat{h}_{i_1 j_1}^{(s_1)}(\vec{k}_1)\hat{h}_{i_2 j_2}^{(s_2)}(\vec{k}_2)\hat{h}_{i_3 j_3}^{(s_3)}(\vec{k}_3)\rangle = i\int_{-\infty}^{\tau_0}\mathrm{d}\tau\langle 0|\left[H_{\text{int}}^{\text{PV}}(\tau), \hat{h}_{i_1 j_1}^{(s_1)}(\vec{k}_1)\hat{h}_{i_2 j_2}^{(s_2)}(\vec{k}_2)\hat{h}_{i_3 j_3}^{(s_3)}(\vec{k}_3)\right]|0\rangle, \tag{3.2}$$

where the interaction Hamiltonian from the cubic Weyl term denoted by $H_{\text{int}}^{\text{PV}}$ is obtained by expanding the cubic Weyl action up to the cubic order in $h_{ij}$ as [59],

$$H_{\text{int}}^{\text{PV}} = -\int d^3x \frac{f(\tau)}{4a^2 M_p^2} \epsilon^{ijk} \Big\{ 4 h'_{pj,k} h'_{pm,l}(h'_{il,m} - h'_{im,l}) \\ + \left(h''_{kq} + \partial^2 h_{kq}\right) \left[ -3 h'_{iq,m} \left(h''_{jm} + \partial^2 h_{jm}\right) + h'_{mq,j}\left(h''_{im} + \partial^2 h_{im}\right)\right] \Big\}. \quad (3.3)$$

As will be summarized in appendix B, by using the explicit form of the polarization tensor, one can simplify eq. (3.2) into the following expression:

$$\langle \hat{h}^{(s_1)}(\vec{k}_1) \hat{h}^{(s_2)}(\vec{k}_2) \hat{h}^{(s_3)}(\vec{k}_3) \rangle \\ = (2\pi)^3 \delta^{(3)}(\vec{k}_1 + \vec{k}_2 + \vec{k}_3) F(s_1 k_1, s_2 k_2, s_3 k_3) \left[ \mathcal{I}_0 + (\mathcal{I}_1 + 2 \text{ permutations})\right], \\ = (2\pi)^7 \delta^{(3)}(\vec{k}_1 + \vec{k}_2 + \vec{k}_3) F(s_1 k_1, s_2 k_2, s_3 k_3) \frac{\mathcal{P}_h^2}{k_1^3 k_2^3 k_3^3} \left[\mathcal{A}_0 + (\mathcal{A}_1 + 2 \text{ permutations})\right], \quad (3.4)$$

where

$$F(x,y,z) \equiv \frac{1}{64 x^2 y^2 z^2}(x+y+z)^3(x-y+z)(x+y-z)(x-y-z), \quad (3.5)$$

$$\mathcal{I}_0 \equiv s_1 s_2 s_3 k_1 k_2 k_3 \text{Im}\left[\int_{-\infty}^{\tau_0} d\tau \frac{12 f(\tau)}{a^2 M_p^2} \left[h_{k_1}^*(\tau_0) h_{k_2}^*(\tau_0) h_{k_3}^*(\tau_0) h'_{k_1}(\tau) h'_{k_2}(\tau) h'_{k_3}(\tau)\right]\right], \quad (3.6)$$

$$\mathcal{I}_1 \equiv s_1 k_1 \text{Im}\left[\int_{-\infty}^{\tau_0} d\tau \frac{12 f(\tau)}{a^2 M_p^2} h_{k_1}^*(\tau_0) h_{k_2}^*(\tau_0) h_{k_3}^*(\tau_0) h'_{k_1}(\tau) \\ \times [\mathcal{H}(\tau) h'_{k_2}(\tau) + k_2^2 h_{k_2}(\tau)][\mathcal{H}(\tau) h'_{k_3}(\tau) + k_3^2 h_{k_3}(\tau)]\right], \quad (3.7)$$

and $\mathcal{A}_\bullet$ is originating from $\mathcal{I}_\bullet$ with $\bullet = 0, 1$. In the present paper, as analogous to scalar perturbations, we introduce $f_{\text{NL}}$ to quantify the amplitude of the tensor non-Gaussianity,

$$f_{\text{NL}} := \frac{\mathcal{A}}{\sum_i k_i^3}, \quad (3.8)$$

and we evaluate this parameter at the squeezed limit ($k_L := k_1 \ll k_2 = k_3 =: k_S$) and the equilateral limit ($k_1 = k_2 = k_3$). The corresponding $f_{\text{NL}}$ evaluated at the squeezed and equilateral limits are denoted by $f_{\text{NL}}^{\text{local}}$ and $f_{\text{NL}}^{\text{eq}}$, respectively.

Here, during the matter-dominated contracting phase, the conventional "slow-roll" parameter $\epsilon = -\dot{H}/H^2$ takes $\epsilon = 3/2$ which is much larger than that in the quasi-de Sitter inflation where $\epsilon \ll 1$. As has been shown in ref. [42], the parity-violating signatures in the three-point function from slow-roll inflation with the Bunch-Davies state are suppressed by $\epsilon$. Therefore, one might expect that those signatures from matter bounce with the Minkowski vacuum state would amplify the non-Gaussianities. However, we will show that this is not the case for the cubic-Weyl term with a non-dynamical coupling (i.e., $\lambda = 0$) and a specific dynamical coupling is necessary for the amplification.

## 3.1 Non-dynamical coupling

We here compute the three-point function for the non-dynamical coupling case, i.e., $f(\phi) = b/M_p^2$. The time integral can be evaluated directly

$$\mathcal{I}_0 = \frac{288 b \sum_i k_i^3}{5 k_1^3 k_2^3 k_3^3 M_p^8 \tau_0^5} s_1 s_2 s_3 k_1 k_2 k_3, \ \mathcal{I}_1 = \frac{3456 b \sum_i k_i^3}{17 k_1^3 k_2^3 k_3^3 M_p^8 \tau_0^7} s_1 k_1, \tag{3.9}$$

which gives

$$\mathcal{A}_0 = \frac{9\pi^4}{40} b \mathcal{P}_h^2 \left(\sum_i k_i^3\right) s_1 s_2 s_3 k_1 k_2 k_3 \tau_0^3, \ \mathcal{A}_1 = \frac{27\pi^4}{34} b \mathcal{P}_h^2 \left(\sum_i k_i^3\right) s_1 k_1 \tau_0, \tag{3.10}$$

respectively. One can see that the leading-order contribution to $f_{\rm NL}$ comes from $\mathcal{A}_1$ and its 2 permutations, and hence is suppressed in proportion to $(-k_i \tau_0) \ll 1$, which makes it difficult to detect the tensor non-Gaussianity originating from the cubic Weyl term. Here, the non-Gaussianities are generally generated after the horizon-cross scale, $-k_i \tau \lesssim 1$. This is because the integrands of the time integrals appearing in the in-in formalism are proportional to the exponential function $e^{-i(k_1+k_2+k_3)\tau}$ which rapidly oscillates on the subhorizon scales, $-k_i \tau \gg 1$, and this rapid oscillation eliminates any contributions from the subhorizon scales to the three-point function. In the present case, the mode function grows after horizon crossing in proportion to $(\tau_0/\tau)^3$ which is much smaller than unity for $|\tau| \gg |\tau_0|$ in the time regime away from the end of the contracting phase, e.g., around the horizon-cross scale. Furthermore, all of the cubic interaction terms from the cubic Weyl involve spatial derivatives, and hence those are suppressed on the superhorizon scales. For these reasons, we need to amplify the integrand after horizon crossing. The dynamical coupling of the form $(\tau/\tau_0)^\lambda$ that we adopt is much larger than unity for $|\tau| \gg |\tau_0|$. Accordingly, that dynamical coupling has the potential to cancel the aforementioned suppression. As we will show below, a positive $\lambda$, especially the case of $\lambda > 15$, can generally enhance the amplitude of the three-point function.

## 3.2 Dynamical coupling

One can straightforwardly compute the three-point function with the dynamical coupling as well. As will be shown in appendix C, $f_{\rm NL}$ is dependent of a power of $(-k_i \tau_0)$ in general, and the resultant $f_{\rm NL}^{\rm local}$ and $f_{\rm NL}^{\rm eq}$ are scale dependent. The only exception is the case of $\lambda = 15$ where $f_{\rm NL}$ takes the following form,

$$f_{\rm NL} \simeq -\frac{27}{16} b \pi^5 \mathcal{P}_h^2 \left(2 \sum_i s_i k_i^3 + 3 s_1 s_2 s_3 k_1 k_2 k_3\right) \frac{1}{\sum_i k_i^3} F(s_i, k_i). \tag{3.11}$$

The non-linearity parameters $f_{\rm NL}^{\rm local}$ and $f_{\rm NL}^{\rm eq}$ then read

$$f_{\rm NL}^{\rm local} \simeq \frac{27 b \pi^5}{512} \mathcal{P}_h^2 \left[8(s_2+s_3) + 3 s_1 s_2 s_3 \left(\frac{k_L}{k_S}\right)^3\right], \tag{3.12}$$

$$f_{\rm NL}^{\rm eq} \simeq \frac{27 b \pi^5}{1024} \mathcal{P}_h^2 \left(20 \sum_i s_i + 21 s_1 s_2 s_3\right), \tag{3.13}$$

the leading-order terms of which are scale independent. Please note here that $f_{\rm NL}$ is anti symmetric under replacements of $s_i$, e.g., $f_{\rm NL}^{\rm local}(s_1=1, s_2=1, s_3=1) = -f_{\rm NL}^{\rm local}(s_1=$ $-1, s_2 = -1, s_3 = -1$), as a consequence of parity violation (see, e.g., ref. [83] for a parity-odd case). In this case, $f_{\mathrm{NL}}$ is not suppressed by positive powers of $(-k\tau_0)$, and the amplitude from matter bounce can be larger than that from slow-roll inflation with the Bunch-Davies vacuum state if we assume that the tensor power spectrum between bounce and inflation are the same order of magnitude. However, the amplitude is still suppressed by $\mathcal{P}_h^2$ that is of $\mathcal{O}(10^{-22})$ for the tensor-to-scalar ratio $r$ of $\mathcal{O}(10^{-2})$ where $r := \mathcal{P}_h/\mathcal{P}_\zeta$ with the scalar power spectrum $\mathcal{P}_\zeta$. This would indicate that there is no chance for us to detect it by actual experiments.

We next consider the case of $|\lambda - 15| \geq 1$. In this case, as shown in appendix C, the non-linearity parameter is proportional to $b\mathcal{P}_h^2(-k_i\tau_0)^n$ with $|n| \geq \mathcal{O}(1)$. Here, for the wavenumber mode $k_{\mathrm{CMB}} = 0.02 \mathrm{Mpc}^{-1}$ which is the pivot scale of Planck, we have

$$-k_{\mathrm{CMB}}\tau_0 = \frac{k_{\mathrm{CMB}}}{M_p} \times \mathcal{O}\left(\frac{M_p}{H_0}\right) = \mathcal{O}\left(10^{-55}r^{-1/2}\right), \tag{3.14}$$

where we used the observed value $\mathcal{P}_\zeta \simeq 2 \times 10^{-9}$. If we take $r = 0.01$ in light of the current constraint $r < 0.056$ [84], then we obtain $-k_{\mathrm{CMB}}\tau_0 = \mathcal{O}(10^{-54})$. Thus, if $f_{\mathrm{NL}}$ has the aforementioned power-law scale dependence, the tensor non-Gaussianity is either overproduced or highly suppressed. The models in the former case are ruled out by current CMB experiments,[3] while those in the latter case can never be tested through the tensor non-Gaussian signatures.

The remaining case is $|\lambda - 15| < 1$. In this case, the resultant non-linearity parameters are proportional to $(-k_i\tau_0)^{\lambda-15}$. Hence, by choosing $\lambda$ appropriately, we can obtain a large $f_{\mathrm{NL}}$ that is consistent with the current constraint while can potentially be tested by upcoming CMB experiments. To clarify the parameter region giving $f_{\mathrm{NL}} \geq \mathcal{O}(1)$, let us estimate $\ln f_{\mathrm{NL}}$ as

$$\ln f_{\mathrm{NL}} \simeq \frac{\lambda - 11}{2}\ln r + (55\lambda - 843)\ln 10, \tag{3.15}$$

where we set $b = 1$, ignored $\mathcal{O}(1)$ coefficients in $f_{\mathrm{NL}}$, and used $\mathcal{P}_\zeta \simeq 2 \times 10^{-9}$. Both $f_{\mathrm{NL}}^{\mathrm{local}}$ and $f_{\mathrm{NL}}^{\mathrm{eq}}$ share the same order of magnitude obtained from eq. (3.15). The plot of $\ln f_{\mathrm{NL}}$ is shown in figure 1. Depending on the value of $\lambda$, we obtain $f_{\mathrm{NL}} \geq \mathcal{O}(1)$ even for $r = 0.01$, e.g., $f_{\mathrm{NL}} = \mathcal{O}(1)$ for $\lambda = 15.4$ and $r = 0.01$. As an example, we also evaluate the shape of the bispectrum for $\lambda = 15.4$ that gives $f_{\mathrm{NL}} = \mathcal{O}(1)$. A plot of $\mathcal{A}_0/k_1k_2k_3$ and that of $\mathcal{A}_1/(k_1k_2k_3)$ are shown in figure 2 and 3, respectively. Those figures show that the bispectrum originating from $\mathcal{A}_0$ and $\mathcal{A}_1$ has a peak at the equilateral and squeezed limit, respectively.

Before closing this subsection, let us comment on strong coupling and classical non-linearity. Once we introduce the dynamical coupling which increases as time goes back, then one may expect that strong coupling occurs in far past (on subhorizon scales). To clarify this point, let us follow refs. [86–92]. In terms of the canonically normalized tensor fluctuation $v_{ij} = aM_p h_{ij}/2$ denoted by $v$ below, one obtains

$$S_v^{(2)} \sim \int \mathrm{d}\tau \mathrm{d}^3 x (\partial_i)^2 v^2, \tag{3.16}$$

$$S_v^{(3)} \sim \int \mathrm{d}\tau \mathrm{d}^3 x \frac{1}{a^5}(-\tau)^\lambda (\partial_i)^6 v^3 =: \int \mathrm{d}\tau \mathrm{d}^3 x \frac{(\partial_i)^6}{\Lambda^5} v^3, \tag{3.17}$$

---

[3]So far, CMB experiments have put constraints on $r^2 f_{\mathrm{NL}}^{\mathrm{local}}$ and $r^2 f_{\mathrm{NL}}^{\mathrm{eq}}$ as $\leq \mathcal{O}(10^3)$ and $\leq \mathcal{O}(10^4)$, respectively [85]. If $f_{\mathrm{NL}}$ from bounce is enhanced in proportion to negative powers of $(-k_i\tau_0) \ll 1$, then those bounce models are ruled out.

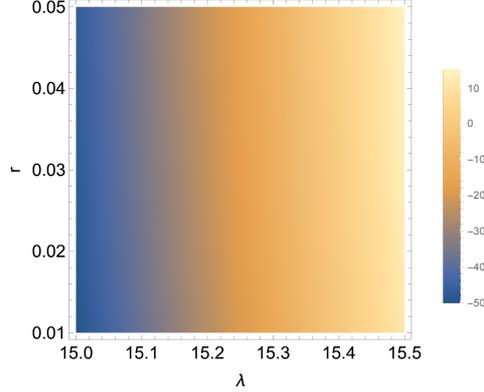

**Figure 1.** A plot of $\ln f_{\rm NL}$ as a function of $r$ and $\lambda$.

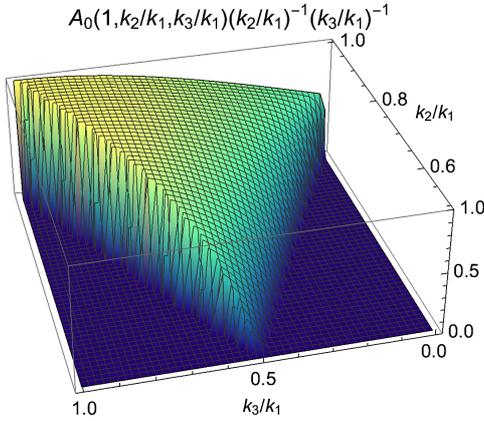

**Figure 2.** A plot of $\mathcal{A}_0/(k_1 k_2 k_3)$ as a function of $k_2/k_1$ and $k_3/k_1$. We normalized it to 1 for the equilateral triangle $k_1 = k_2 = k_3$.

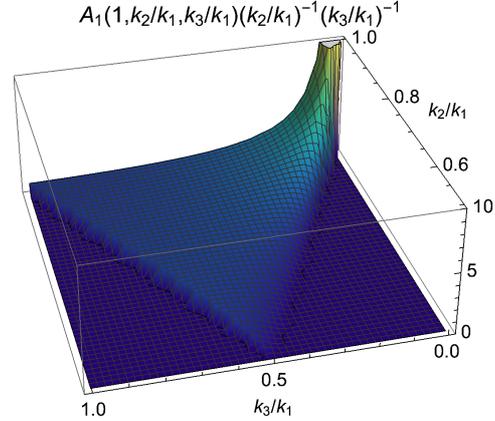

**Figure 3.** A plot of $\mathcal{A}_1/(k_1 k_2 k_3)$ as a function of $k_2/k_1$ and $k_3/k_1$. We normalized it to 1 for the equilateral triangle $k_1 = k_2 = k_3$.

where $S_v^{(2)}$ and $S_v^{(3)}$ stand for the quadratic and cubic actions for the canonically normalized perturbations, respectively. We also defined $\Lambda := a(-\tau)^{-\lambda/5}$. Since we have the following time dependence

$$\Lambda \propto (-\tau)^{(10-\lambda)/5}, \tag{3.18}$$

which is asymptotic to 0 in the past infinity for $\lambda > 10$, strong coupling can occur for that case. Here, the characteristic classical energy scale of the contracting background is the Hubble parameter which also approaches 0 as time goes back. We thus require that the strong coupling scale $\Lambda$ is much higher than the classical energy scale of the contracting spacetime in the past infinity:

$$H \propto \tau^{-3} \ll \Lambda \propto (-\tau)^{(10-\lambda)/5}, \tag{3.19}$$

to evade strong coupling. As a result, the condition to avoid strong coupling reads $\lambda < 25$ that includes the case of $|\lambda - 15| < 1$.

Also, after the perturbations cross the horizon, classical non-linearity may cause breakdown of the linear perturbation theory. The non-linear correction to the linear perturbation would be of $\mathcal{O}(f_{\rm NL}|h_{ij}|)$. Here, the enhanced $f_{\rm NL}$ which is still allowed by the current CMB experiments has the scale dependence of $(-k_i\tau_0)^{15-\lambda}$ with $\lambda > 15$. In this case, the possible maximum enhancement of $f_{\rm NL}$ is obtained at the largest scale, i.e., $k = k_{\rm CMB} \sim 10^{-2}{\rm Mpc}^{-1}$. On these scales, we have $|h_{ij}| = \mathcal{O}(\mathcal{P}_h^{1/2}) \leq 10^{-5}$ where we used $r \leq \mathcal{O}(10^{-2})$. Thus, as long as $f_{\rm NL}(k_{\rm CMB}) < \mathcal{O}(10^5)$, the tensor perturbations are in the linear perturbation regime. For instance, we obtain $f_{\rm NL}(k_{\rm CMB}) = \mathcal{O}(10^5)$ for $r = 0.01$ and $\lambda \simeq 15.5$. Therefore, we conclude this section as there is indeed a parameter region where the parity-violating signatures in the tensor non-Gaussianities can be enhanced up to $f_{\rm NL} \leq \mathcal{O}(10^5)$ within the perturbative regime.

## 4 Conclusion and outlook

In this paper, we have investigated the parity-violating signatures in the primordial tensor bispectrum from the cubic Weyl term in matter bounce cosmology. The parity-violating signatures have been explored at the three-point function level for the first time in that context in this paper. First, we have presented the primordial tensor bispectrum with the non-dynamical coupling $\lambda = 0$. Although there is no slow-roll suppression in contrast to inflation as expected, we have shown that the non-linearity parameter is scale dependent and highly suppressed compared to that from inflation. To enhance the non-Gaussianities, we have introduced the dynamical coupling of the form, $f(\phi) \propto (\tau/\tau_0)^\lambda$, that can boost the magnitudes of the cubic interactions. For the dynamical coupling case, the non-linearity parameter generally has the scale dependence leading to the overproduction or suppression of the non-Gaussianities, which is either ruled out or never be tested by CMB experiments, respectively. However, the case of $|\lambda - 15| < 1$ is the exception, and the primordial bispectrum can potentially be tested by CMB experiments. Here, we have found that a non-exact-de Sitter background is not enough to obtain sizable tensor non-Gaussianities from the cubic Weyl action in the context of matter bounce cosmology. We have also investigated the conditions to avoid strong coupling problems and confirmed that the parameter space for $|\lambda - 15| < 1$ does not suffer from those.

Here, as has been investigated in refs. [66, 93], the three-point function of the tensor perturbations originating from the Einstein-Hilbert action in matter bounce cosmology is highly suppressed so that it is difficult to detect the non-Gaussian signatures by CMB experiments. We thus emphasize that, similarly to the case of inflation [59], the cubic Weyl term with the dynamical coupling has the potential interest in looking for the early universe models that could actually be tested through the primordial tensor non-Gaussianities in the context of alternatives to inflation. Hence, it would be important to investigate the impacts of the enhanced tensor non-Gaussianities on CMB bispectra. We will leave it to the future work. Also, we have focused on the contracting phase, but depending on the models, the subsequent phases can leave impacts on the observational signatures (see, e.g., ref. [94] for the impact of the bouncing phase on the primordial tensor power spectrum in dynamical Chern-Simons gravity). Therefore, it would be interesting to consider the subsequent phases in explicit models. It would also be worth investigating the parity-violating correlation functions from different non-singular scenarios, e.g., Ekpyrotic cosmology [95–97] and Galilean Genesis [98, 99]. Non-canonical inflationary scenarios [100–102] can also predict sizable parity-violating primordial fluctuations [37, 38]. A comparison of parity-violating non-Gaussianities between inflation and non-singular cosmology can potentially distinguish those scenarios by experiments. As a further extension of the present work, cross-correlation three-point functions originating from scalar-scalar-tensor and scalar-tensor-tensor interactions and higher-order correlation functions (e.g., primordial trispectra) would also be important as well as the above.

## Acknowledgments

We thank Yong Cai, Shin'ichi Hirano, Chunshan Lin, and Giorgio Orlando for fruitful discussions and useful comments on the manuscript. The work of S.A. was supported by the grant No. UMO-2021/42/E/ST9/00260 from the National Science Centre, Poland, and MEXT-JSPS Grant-in-Aid for Transformative Research Areas (A) "Extreme Universe", No. JP21H05189. The work of M.Z. was supported by grant No. UMO 2021/42/E/ST9/00260 from the National Science Centre, Poland.

## A  Polarization tensor

In this section, we fix the representation of polarization tensors following the convention in [103]. The momentum conservation $\sum_{i=1}^{3} \vec{k}_i = 0$ enables us to set all $\vec{k}_i$'s in a plane without loss of generality. We choose $\vec{k}_1$ to be in $x$-direction, and all $\vec{k}_i$'s are in $(x, y)$ plane, and hence we have

$$\vec{k}_1 = k_1(1, 0, 0), \ \vec{k}_2 = k_2(\cos\theta, \sin\theta, 0), \ \vec{k}_3 = k_3(\cos\phi, \sin\phi, 0), \tag{A.1}$$

where

$$\cos\theta = \frac{k_3^2 - k_1^2 - k_2^2}{2k_1 k_2}, \ \sin\theta = \frac{\sigma}{2k_1 k_2},$$
$$\cos\phi = \frac{k_2^2 - k_1^2 - k_3^2}{2k_1 k_3}, \ \sin\phi = -\frac{\sigma}{2k_1 k_3}, \tag{A.2}$$

with

$$\sigma \equiv \sqrt{\sum_{i \neq j} k_i^2 k_j^2 - \sum_i k_i^4}, \tag{A.3}$$

and $\theta \in [0, \pi]$ and $\phi \in [\pi, 2\pi]$. The polarization tensor for $\vec{k}_1$ using the representation (A.1) is simply

$$e^{(s_1)}(\vec{k}_1) = \frac{1}{2} \begin{pmatrix} 0 & 0 & 0 \\ 0 & 1 & is_1 \\ 0 & -is_1 & -1 \end{pmatrix}. \tag{A.4}$$

The other two polarization tensors can be obtained by rotating $e^{(s_1)}(\vec{k}_1)$ by the angle $\theta$ and $\phi$ respectively:

$$e^{(s_2)}(\vec{k}_2) = \frac{1}{2} \begin{pmatrix} \sin^2\theta & -\sin\theta\cos\theta & -is_2\sin\theta \\ -\sin\theta\cos\theta & \cos^2\theta & is_2\cos\theta \\ -is_2\sin\theta & is_2\cos\theta & -1 \end{pmatrix}, \tag{A.5}$$

$$e^{(s_3)}(\vec{k}_3) = \frac{1}{2}\begin{pmatrix} \sin^2\phi & -\sin\phi\cos\phi & -is_3\sin\phi \\ -\sin\phi\cos\phi & \cos^2\phi & is_3\cos\phi \\ -is_3\sin\phi & is_3\cos\phi & -1 \end{pmatrix}. \tag{A.6}$$

## B  Computation of tensor bispectra

We present the computation of tensor bispectra in this appendix. We start by evaluating the following term:

$$\begin{aligned}
\langle 0|H_{\rm int}\hat{h}^{(s_1)}_{i_1j_1}(\vec{k}_1)\hat{h}^{(s_2)}_{i_2j_2}(\vec{k}_2)\hat{h}^{(s_3)}_{i_3j_3}(\vec{k}_3)|0\rangle &= -\frac{if(\tau)}{4a^2M_p^2}\epsilon^{ijk}\sum_{s_as_bs_c}\int d^3p_1 d^3p_2 d^3p_3 \delta(\vec{p}_1+\vec{p}_2+\vec{p}_3) \\
&\times \langle 0|\hat{a}_{s_a}(\vec{p}_1)\hat{a}_{s_b}(\vec{p}_2)\hat{a}_{s_c}(\vec{p}_3)\hat{a}^\dagger_{s_1}(-\vec{k}_1)\hat{a}^\dagger_{s_2}(-\vec{k}_2)\hat{a}^\dagger_{s_3}(-\vec{k}_3)|0\rangle h^*_{k_1}h^*_{k_2}h^*_{k_3}e^{(s_1)*}_{i_1j_1}(-\vec{k}_1)e^{(s_2)*}_{i_2j_2}(-\vec{k}_2)e^{(s_3)*}_{i_3j_3}(-\vec{k}_3) \\
&\times \Big[(h''_{p_1}-p_1^2 h_{p_1})e^{(s_a)}_{kq}(\vec{p}_1)(h''_{p_2}-p_2^2 h_{p_2})h'_{p_3}\Big(p_{3,j}e^{(s_b)}_{im}(\vec{p}_2)e^{(s_c)}_{mq}(\vec{p}_3)-3p_{3,m}e^{(s_b)}_{jm}(\vec{p}_2)e^{(s_c)}_{iq}(\vec{p}_3)\Big) \\
&-4p_{1,k}p_{2,l}h'_{p_1}h'_{p_2}h'_{p_3}e^{(s_a)}_{pj}(\vec{p}_1)e^{(s_b)}_{pm}(\vec{p}_2)\Big(p_{3,m}e^{(s_c)}_{il}(\vec{p}_3)-(l\longleftrightarrow m)\Big)\Big]\ (+5\text{ perm.}) \\
&= -\delta^{(3)}(\vec{k}_1+\vec{k}_2+\vec{k}_3)\epsilon^{ijk}\frac{if(\tau)}{a^2M_p^2}h^*_{k_1}h^*_{k_2}h^*_{k_3}e^{(s_1)}_{i_1j_1}(\vec{k}_1)e^{(s_2)}_{i_2j_2}(\vec{k}_2)e^{(s_3)}_{i_3j_3}(\vec{k}_3)h'_{k_3} \\
&\times \Big[(\mathcal{H}h'_{k_1}+k_1^2 h_{k_1})(\mathcal{H}h'_{k_2}+k_2^2 h_{k_2})e^{(s_1)*}_{kq}(\vec{k}_1)\Big(k_{3,j}e^{(s_2)*}_{im}(\vec{k}_2)e^{(s_3)*}_{mq}(\vec{k}_3)-3k_{3,m}e^{(s_2)*}_{jm}(\vec{k}_2)e^{(s_3)*}_{iq}(\vec{k}_3)\Big) \\
&-k_{1,k}k_{2,l}h'_{k_1}h'_{k_2}e^{(s_1)*}_{pj}(\vec{k}_1)e^{(s_2)*}_{pm}(\vec{k}_2)\Big(k_{3,m}e^{(s_3)*}_{il}(\vec{k}_3)-(l\longleftrightarrow m)\Big)\Big]\ (+5\text{ perm.}).
\end{aligned} \tag{B.1}$$

Here, in the exact-de Sitter spacetime, the mode function is of the form $h_k(\tau) \propto e^{-ik\tau}(1+ik\tau)$, which yields $\mathcal{H}h' + k^2 h_k = ikh_k$, and then the above expression simplifies to

$$\begin{aligned}
\delta^{(3)}(\vec{k}_1+\vec{k}_2+\vec{k}_3)\frac{if(\tau)}{a^2M_p^2}\epsilon^{ijk} h'_{k_1}h'_{k_2}h'_{k_3}h^*_{k_1}h^*_{k_2}h^*_{k_3}e^{(s_1)}_{i_1j_1}(\vec{k}_1)e^{(s_2)}_{i_2j_2}(\vec{k}_2)e^{(s_3)}_{i_3j_3}(\vec{k}_3) \\
\times \Big[k_1 k_2 e^{(s_1)*}_{kq}(\vec{k}_1)\Big(k_{3,j}e^{(s_2)*}_{im}(\vec{k}_2)e^{(s_3)*}_{mq}(\vec{k}_3)-3k_{3,m}e^{(s_2)*}_{jm}(\vec{k}_2)e^{(s_3)*}_{iq}(\vec{k}_3)\Big) \\
+k_{1,k}k_{2,l}e^{(s_1)*}_{pj}(\vec{k}_1)e^{(s_2)*}_{pm}(\vec{k}_2)\Big(k_{3,m}e^{(s_3)*}_{il}(\vec{k}_3)-(l\longleftrightarrow m)\Big)\Big]
\end{aligned}$$

which reproduces the result in [59]. In the matter-dominated contracting universe, the simple relation does not hold. The final expression of (B.1) takes the following form,

$$\begin{aligned}
\langle 0|H_{\rm int}\hat{h}^{(s_1)}_{i_1j_1}(\vec{k}_1)\hat{h}^{(s_2)}_{i_2j_2}(\vec{k}_2)\hat{h}^{(s_3)}_{i_3j_3}(\vec{k}_3)|0\rangle &= -\delta^{(3)}(\vec{k}_1+\vec{k}_2+\vec{k}_3)\epsilon^{ijk}\frac{if(\tau)}{a^2M_p^2}h^*_{k_1}h^*_{k_2}h^*_{k_3}h'_{k_3} \\
&\times \Big[k_{1,k}k_{2,l}h'_{k_1}h'_{k_2}\Pi^{(s_1)}_{i_1j_1,pj}(\vec{k}_1)\Pi^{(s_2)}_{i_2j_2,pm}(\vec{k}_2)\Big(k_{3,m}\Pi^{(s_3)}_{i_3j_3,il}(\vec{k}_3)-(l\longleftrightarrow m)\Big) \\
&-(\mathcal{H}h'_{k_1}+k_1^2 h_{k_1})(\mathcal{H}h'_{k_2}+k_2^2 h_{k_2})\Pi^{(s_1)}_{i_1j_1,kq}(\vec{k}_1) \\
&\times \Big(k_{3,j}\Pi^{(s_2)}_{i_2j_2,im}(\vec{k}_2)\Pi^{(s_3)}_{i_3j_3,mq}(\vec{k}_3)-3k_{3,m}\Pi^{(s_2)}_{i_2j_2,jm}(\vec{k}_2)\Pi^{(s_3)}_{i_3j_3,iq}(\vec{k}_3)\Big)\Big],
\end{aligned} \tag{B.2}$$

where we introduced

$$\Pi^{(s)}_{ij,kl}(\vec{p}) \equiv e^{(s)}_{ij}(\vec{p})e^{(s)*}_{kl}(\vec{p}), \tag{B.3}$$

which satisfies

$$\Pi^{(s)}_{ij,kl}(\vec{p}) = \Pi^{(-s)}_{kl,ij}(\vec{p}),\ \ \Pi^{(s)}_{ij,kl}(\vec{p})e^{(s')*}_{ij}(\vec{p}) = \delta_{ss'}e^{(s)*}_{kl}(\vec{p}). \tag{B.4}$$

Now we can sum up the correlation function with polarization tensors. After making use of (B.4) we arrive

$$
\begin{aligned}
\langle 0|H_{\text{int}} &\hat{h}_{i_1j_1}^{(s_1)}(\vec{k}_1)\hat{h}_{i_2j_2}^{(s_2)}(\vec{k}_2)\hat{h}_{i_3j_3}^{(s_3)}(\vec{k}_3)|0\rangle e_{i_1j_1}^{(s_1)*}(\vec{k}_1)e_{i_2j_2}^{(s_2)*}(\vec{k}_2)e_{i_3j_3}^{(s_3)*}(\vec{k}_3) \\
&= -\delta^{(3)}(\vec{k}_1+\vec{k}_2+\vec{k}_3)\epsilon^{ijk}\frac{if(\tau)}{a^2 M_p^2}h_{k_1}^*(\tau_0)h_{k_2}^*(\tau_0)h_{k_3}^*(\tau_0)h_{k_3}'(\tau) \\
&\quad \times \Big[k_{1,k}k_{2,l}h_{k_1}'(\tau)h_{k_2}'(\tau)e_{pj}^{(s_1)*}(\vec{k}_1)e_{pm}^{(s_2)*}(\vec{k}_2)\left(k_{3,m}e_{il}^{(s_3)*}(\vec{k}_3)-(l\longleftrightarrow m)\right) \\
&\quad -[\mathcal{H}(\tau)h_{k_1}'(\tau)+k_1^2 h_{k_1}(\tau)][\mathcal{H}(\tau)h_{k_2}'(\tau)+k_2^2 h_{k_2}(\tau)]e_{kq}^{(s_1)*}(\vec{k}_1) \\
&\quad \times \left(k_{3,j}e_{im}^{(s_2)*}(\vec{k}_2)e_{mq}^{(s_3)*}(\vec{k}_3)-3k_{3,m}e_{jm}^{(s_2)*}(\vec{k}_2)e_{iq}^{(s_3)*}(\vec{k}_3)\right)\Big].
\end{aligned}
\quad \text{(B.5)}
$$

By adding the complex conjugate of the above, we obtain

$$
\begin{aligned}
\langle \hat{h}_{i_1j_1}^{(s_1)}(\vec{k}_1)\hat{h}_{i_2j_2}^{(s_2)}(\vec{k}_2)\hat{h}_{i_3j_3}^{(s_3)}(\vec{k}_3)\rangle &= \delta^{(3)}(\vec{k}_1+\vec{k}_2+\vec{k}_3)\int_{-\infty}^{\tau_0}d\tau\frac{2if(\tau)}{a^2 M_p^2}\epsilon^{ijk} \\
&\quad \times \Big\{\text{Im}\left[h_{k_1}^*(\tau_0)h_{k_2}^*(\tau_0)h_{k_3}^*(\tau_0)h_{k_1}'(\tau)h_{k_2}'(\tau)h_{k_3}'(\tau)\right] \\
&\quad \times k_{1,k}k_{2,l}e_{pj}^{(s_1)*}(\vec{k}_1)e_{pm}^{(s_2)*}(\vec{k}_2)\left(k_{3,m}e_{il}^{(s_3)*}(\vec{k}_3)-(l\longleftrightarrow m)\right) \\
&\quad -\text{Im}\left[h_{k_1}^*(\tau_0)h_{k_2}^*(\tau_0)h_{k_3}^*(\tau_0)[\mathcal{H}(\tau)h_{k_1}'(\tau)+k_1^2 h_{k_1}(\tau)][\mathcal{H}(\tau)h_{k_2}'(\tau)+k_2^2 h_{k_2}(\tau)]h_{k_3}'(\tau)\right] \\
&\quad \times e_{kq}^{(s_1)*}(\vec{k}_1)\left(k_{3,j}e_{im}^{(s_2)*}(\vec{k}_2)e_{mq}^{(s_3)*}(\vec{k}_3)-3k_{3,m}e_{jm}^{(s_2)*}(\vec{k}_2)e_{iq}^{(s_3)*}(\vec{k}_3)\right)\Big\} + (5\text{ perm.}).
\end{aligned}
\quad \text{(B.6)}
$$

The correlation function (B.6) can be simplified by using the expressions of polarization tensors in a specific basis from appendix A. For instance,

$$
\begin{aligned}
&\epsilon^{ijk}k_{1,k}k_{2,l}e_{pj}^{(s_1)*}(\vec{k}_1)e_{pm}^{(s_2)*}(\vec{k}_2)\left(k_{3,m}e_{il}^{(s_3)*}(\vec{k}_3)-(l\longleftrightarrow m)\right) + (5\text{ perm.}) \\
&= -6is_1 s_2 s_3 k_1 k_2 k_3 F(s_1 k_1, s_2 k_2, s_3 k_3).
\end{aligned}
\quad \text{(B.7)}
$$

For the last two lines of (B.6), since the mode function is symmetric over $h_1$ and $h_2$, but not $h_3$, we can only sum over the $1\longleftrightarrow 2$ permutation:

$$
\begin{aligned}
&\epsilon^{ijk}e_{kq}^{(s_1)*}(\vec{k}_1)\left(k_{3,j}e_{im}^{(s_2)*}(\vec{k}_2)e_{mq}^{(s_3)*}(\vec{k}_3)-3k_{3,m}e_{jm}^{(s_2)*}(\vec{k}_2)e_{iq}^{(s_3)*}(\vec{k}_3)\right) + \left(s_1,\vec{k}_1\longleftrightarrow s_2,\vec{k}_2\right) \\
&= -6iFk_3 s_3.
\end{aligned}
\quad \text{(B.8)}
$$

Thus the correlation function (B.6) can be organized in the following form:

$$
\langle \hat{h}^{(s_1)}(\vec{k}_1)\hat{h}^{(s_2)}(\vec{k}_2)\hat{h}^{(s_3)}(\vec{k}_3)\rangle = (2\pi)^3\delta(\vec{k}_1+\vec{k}_2+\vec{k}_3)F(s_1 k_1, s_2 k_2, s_3 k_3)\left(\mathcal{I}_0 + \sum_j \mathcal{I}_j\right), \quad \text{(B.9)}
$$

where

$$
\mathcal{I}_0 \equiv s_1 s_2 s_3 k_1 k_2 k_3 \text{Im}\left[\int_{-\infty}^{\tau_0}d\tau\frac{12f(\tau)}{a^2 M_p^2}\left[h_{k_1}^*(\tau_0)h_{k_2}^*(\tau_0)h_{k_3}^*(\tau_0)h_{k_1}'(\tau)h_{k_2}'(\tau)h_{k_3}'(\tau)\right]\right], \quad \text{(B.10)}
$$

$$
\begin{aligned}
\mathcal{I}_j \equiv s_j k_j \text{Im}\Big[&\int_{-\infty}^{\tau_0}d\tau\frac{12f(\tau)}{a^2 M_p^2}h_{k_1}^*(\tau_0)h_{k_2}^*(\tau_0)h_{k_3}^*(\tau_0)h_{k_j}'(\tau) \\
&\times [\mathcal{H}(\tau)h_{k_{j+1}}'(\tau)+k_{j+1}^2 h_{k_{j+1}}(\tau)][\mathcal{H}(\tau)h_{k_{j+2}}'(\tau)+k_{j+2}^2 h_{k_{j+2}}(\tau)]\Big],
\end{aligned}
\quad \text{(B.11)}
$$

with $j$ being defined modulo 3.

– 13 –

## C  Generic expression of the three-point function for arbitrary $\lambda$

In this section, we show the results of the three-point function for arbitrary $\lambda$. Introducing the following dimensionless quantities,

$$u \equiv K\tau_0\,,\ v \equiv K\tau\,,\ x_i \equiv k_i/K < 1;\ i = 1, 2, 3\,, \tag{C.1}$$

where $K := k_1 + k_2 + k_3$, one can rewrite the time integrals in eq. (B.10) and eq. (B.11) as

$$\mathcal{I}_0 = s_1 s_2 s_3 K^2 \text{Im} \int_{-\infty}^{u} dv \frac{96 b e^{i(u-v)}}{x_1^2 x_2^2 x_3^2 v^9 M_p^8} \left(\frac{v}{u}\right)^{\lambda-7} \left[\Pi_{j=1}^{3}(1 - i x_j u)\right] \sum_{l=0}^{6} \mathcal{P}_{0,l}(x_1, x_2, x_3) i^{2+l} v^l\,, \tag{C.2}$$

$$\mathcal{I}_1 = s_1 x_1 K^2 \text{Im} \int_{-\infty}^{u} dv \frac{96 b e^{i(u-v)}}{x_1^3 x_2^3 x_3^3 v^{11} M_p^8} \left(\frac{v}{u}\right)^{\lambda-7} \left[\Pi_{j=1}^{3}(1 - i x_j u)\right] \sum_{l=0}^{8} \mathcal{P}_{1,l}(x_1, x_2, x_3) i^{l+2} v^l\,, \tag{C.3}$$

where

$$\mathcal{P}_{0,0} = 27, \quad \mathcal{P}_{0,1} = 27, \quad \mathcal{P}_{0,2} = 9 \sum_i x_i^2 + 27 \sum_{i<j} x_i x_j, \quad \mathcal{P}_{0,3} = 9 \sum_{i<j} x_i x_j, \tag{C.4}$$

$$\mathcal{P}_{0,4} = 3 \sum_{i<j} x_i^2 x_j^2 + 9 x_1 x_2 x_3, \quad \mathcal{P}_{0,5} = 3 x_1 x_2 x_3 \sum_{i<j} x_i x_j, \quad \mathcal{P}_{0,6} = x_1^2 x_2^2 x_3^2, \tag{C.5}$$

$$\mathcal{P}_{1,0} = 108,\ \mathcal{P}_{1,1} = 108, \quad \mathcal{P}_{1,2} = 18(3 - x_1^2), \quad \mathcal{P}_{1,3} = 18(1 - x_1^2),$$
$$\mathcal{P}_{1,4} = 18(1 - x_1)^2(1 - x_2)(1 - x_3) - 9 x_2^2 x_3^2, \tag{C.6}$$
$$\mathcal{P}_{1,5} = 3[3 x_2^2 x_3^2 (1 - x_1) + 2 x_1^2 (1 - x_1)^3 + 3 x_1 x_2 x_3 (2 x_2^2 + 3 x_2 x_3 + 2 x_3^2)], \tag{C.7}$$
$$\mathcal{P}_{1,6} = 3 x_2 x_3 [x_2^2 x_3^2 + 3 x_1 x_2 x_3 + 2 x_1^2 (x_2^2 + x_3^2)], \quad \mathcal{P}_{1,7} = 3 x_1 x_2^2 x_3^2 (x_1 x_2 + x_2 x_3 + x_3 x_1),$$
$$\mathcal{P}_{1,8} = x_1^2 x_2^2 x_3^2. \tag{C.8}$$

Based on the above, we evaluate

$$\mathcal{A}_0 = \frac{k_1^3 k_2^3 k_3^3}{(2\pi)^4 \mathcal{P}_h^2} \mathcal{I}_0,\ \mathcal{A}_1 = \frac{k_1^3 k_2^3 k_3^3}{(2\pi)^4 \mathcal{P}_h^2} \mathcal{I}_1. \tag{C.9}$$

After using the following recurrence formula,

$$\Gamma(\lambda + 1, -i|u|) = \lambda \Gamma(\lambda, -i|u|) + e^{-iu}(-i|u|)^\lambda, \tag{C.10}$$

we obtain

$$\mathcal{A}_0 = \frac{3 K^7}{2 M_p^4} b s_1 s_2 s_3 x_1 x_2 x_3 \left[\frac{\mathcal{B}_0}{|u|} + \frac{\mathcal{B}_1}{|u|^3} \text{Re}[\mathcal{Q}_0(|u|)]\right]$$
$$= K^3 \frac{3}{2 M_p^4 \tau_0^4} b s_1 s_2 s_3 x_1 x_2 x_3 \Big[|u|^3 \mathcal{B}_0 + |u| \mathcal{B}_1 \text{Re}[\mathcal{Q}_0(|u|)]\Big], \tag{C.11}$$

$$\mathcal{A}_1 = \frac{3 K^7}{2 M_p^4} b s_1 x_1 \left[\frac{\mathcal{C}_0}{|u|} + \frac{\mathcal{C}_1}{|u|^3} + \frac{\mathcal{C}_2}{|u|^5} \text{Re}[\mathcal{S}_0(|u|)]\right]$$
$$= K^3 \frac{3}{2 M_p^4 \tau_0^4} b s_1 x_1 \Big[|u|^3 \mathcal{C}_0 + |u| \mathcal{C}_1 + \frac{\mathcal{C}_2}{|u|} \text{Re}[\mathcal{S}_0(|u|)]\Big], \tag{C.12}$$

where

$$\mathcal{B}_0 := x_1 x_2 x_3 \mathcal{P}_{0,1} - \left(\sum_{i<j} x_i x_j + (\lambda - 14) x_1 x_2 x_3\right) \mathcal{P}_{0,2} + \left(1 + (\lambda - 13) S\right) \mathcal{P}_{0,3}$$
$$+ (\lambda - 13)(1 + (\lambda - 12) S) \mathcal{P}_{0,4} + (\lambda - 13)(\lambda - 11)(1 + (\lambda - 12) S) \mathcal{P}_{0,5}$$
$$+ (\lambda - 13)(\lambda - 11)(\lambda - 10)(1 + (\lambda - 12) S) \mathcal{P}_{0,6}, \tag{C.13}$$

$$\mathcal{B}_1 := -\mathcal{P}_{0,1} - (\lambda - 15)\mathcal{P}_{0,2} - (\lambda - 15)(\lambda - 13)\mathcal{P}_{0,3} - (\lambda - 15)(\lambda - 13)(\lambda - 12)\mathcal{P}_{0,4}$$
$$- (\lambda - 15)(\lambda - 13)(\lambda - 12)(\lambda - 11)\mathcal{P}_{0,5} - (\lambda - 15)(\lambda - 13)(\lambda - 12)(\lambda - 11)(\lambda - 10)\mathcal{P}_{0,6},$$
$$\tag{C.14}$$

$$\mathcal{Q}_0 := 1 + (\lambda - 14) e^{-i|u|} u^{14} (-i|u|)^{-\lambda} \left[\Pi_{j=1}^3 (1 + i x_j |u|)\right] \Gamma[\lambda - 15, -i|u|], \tag{C.15}$$

$$\mathcal{C}_0 := -\Big[x_1 x_2 x_3 \mathcal{P}_{1,3} + S \mathcal{P}_{1,4} + (1 + (\lambda - 13) S) \mathcal{P}_{1,5} + (\lambda - 13)(1 + (\lambda - 12) S) \mathcal{P}_{1,6}$$
$$+ (\lambda - 13)(\lambda - 11)(1 + (\lambda - 12) S) \mathcal{P}_{1,7} + (\lambda - 13)(\lambda - 11)(\lambda - 10)(1 + (\lambda - 12) S) \mathcal{P}_{1,8}\Big],$$
$$\tag{C.16}$$

$$\mathcal{C}_1 := x_1 x_2 x_3 \mathcal{P}_{1,1} + D \mathcal{P}_{1,2} + [1 + D(\lambda - 15)] \mathcal{P}_{1,3} + (\lambda - 15)[1 + D(\lambda - 14)] \mathcal{P}_{1,4}$$
$$+ (\lambda - 13)(\lambda - 15)[1 + D(\lambda - 14)] \mathcal{P}_{1,5} + (\lambda - 12)(\lambda - 13)(\lambda - 15)[1 + D(\lambda - 14)] \mathcal{P}_{1,6}$$
$$+ (\lambda - 11)(\lambda - 12)(\lambda - 13)(\lambda - 15)[1 + D(\lambda - 14)] \mathcal{P}_{1,7}$$
$$+ (\lambda - 10)(\lambda - 11)(\lambda - 12)(\lambda - 13)(\lambda - 15)[1 + D(\lambda - 15)] \mathcal{P}_{1,8}, \tag{C.17}$$

$$\mathcal{S}_0 := 1 - (\lambda - 16) e^{-i|u|} |u|^{16} (-i|u|)^{-\lambda} \left[\Pi_{j=1}^3 (1 + i x_j |u|)\right] \Gamma(\lambda - 17, -i|u|), \tag{C.18}$$

$$\mathcal{C}_2 := -\mathcal{P}_{1,1} - (\lambda - 17)\mathcal{P}_{1,2} - (\lambda - 17)(\lambda - 15)\mathcal{P}_{1,3} - (\lambda - 17)(\lambda - 15)(\lambda - 14)\mathcal{P}_{1,4}$$
$$- (\lambda - 17)(\lambda - 15)(\lambda - 14)(\lambda - 13)\mathcal{P}_{1,5} - (\lambda - 17)(\lambda - 15)(\lambda - 14)(\lambda - 13)(\lambda - 12)\mathcal{P}_{1,6}$$
$$- (\lambda - 17)(\lambda - 15)(\lambda - 14)(\lambda - 13)(\lambda - 12)(\lambda - 11)\mathcal{P}_{1,7}$$
$$- (\lambda - 17)(\lambda - 15)(\lambda - 14)(\lambda - 13)(\lambda - 12)(\lambda - 11)(\lambda - 10)\mathcal{P}_{1,8}, \tag{C.19}$$

with

$$S := -x_1 x_2 - x_2 x_3 - x_3 x_1 + (\lambda - 14) x_1 x_2 x_3, \tag{C.20}$$
$$D := -x_1 x_2 - x_2 x_3 - x_3 x_1 + (\lambda - 16) x_1 x_2 x_3. \tag{C.21}$$

On the superhorizon scales (i.e., $|u| \ll 1$), the leading-order contributions to $\text{Re}[\mathcal{Q}_0]$ ($\text{Re}[\mathcal{S}_0]$) are those to $\mathcal{Q}_{\lambda \geq 12}$ and $\mathcal{Q}_{\lambda < 12}$ ($\mathcal{S}_{\lambda \geq 14}$ and $\mathcal{S}_{\lambda < 14}$) defined by

$$\mathcal{Q}_{\lambda \geq 12} := \frac{1}{6}(\lambda - 14)|u|^{14-\lambda}\left[6\cos\left(\frac{\pi\lambda}{2}\right) + 2\left(1 - 3\sum_{i<j} x_i x_j + 3 x_1 x_2 x_3\right)\sin\left(\frac{\pi\lambda}{2}\right)|u|^3\right]\Gamma(\lambda - 15)$$
$$+ \frac{1 + (\lambda - 12) S}{(\lambda - 15)(\lambda - 12)}|u|^2 - \frac{1}{6(\lambda - 13)}\left(1 - 3\sum_{i<j} x_i x_j + 3 x_1 x_2 x_3\right)|u|^4, \tag{C.22}$$

$$\mathcal{Q}_{\lambda < 12} := \frac{1 + (\lambda - 12) S}{(\lambda - 15)(\lambda - 12)}|u|^2, \tag{C.23}$$

$$\mathcal{S}_{\lambda \geq 14} := \frac{1}{6}(16-\lambda)|u|^{16-\lambda}\left[6\cos\left(\frac{\pi\lambda}{2}\right) + 2\left(1 - 3\sum_{i<j} x_i x_j + 3x_1 x_2 x_3\right)\sin\left(\frac{\pi\lambda}{2}\right)|u|^3\right]\Gamma(\lambda-17)$$
$$+ \frac{1+(\lambda-14)D}{(\lambda-17)(\lambda-14)}|u|^2 + \frac{1}{6(\lambda-15)}\left(1 - 3\sum_{i<j} x_i x_j + 3x_1 x_2 x_3\right)|u|^4, \quad \text{(C.24)}$$

$$\mathcal{S}_{\lambda<14} := \frac{1+(\lambda-14)D}{(\lambda-17)(\lambda-14)}|u|^2, \quad \text{(C.25)}$$

where the subscript $\bullet$ on $\mathcal{Q}_\bullet$ and $\mathcal{S}_\bullet$ stand for the conditions under which those four functions are defined. For instance, $\mathcal{Q}_{\lambda\geq 12}$ are defined only for $\lambda \geq 12$. To obtain the leading-order contributions to $\text{Re}[\mathcal{Q}_0]$ and $\text{Re}[\mathcal{S}_0]$, one needs to take the limit of a specific $\lambda$ to $\mathcal{Q}_\bullet$ and $\mathcal{S}_\bullet$ instead of substituting it into those.

One can find from eqs. (C.22)–(C.24) that the non-linearity parameter $f_{\text{NL}} = \mathcal{A}/(\sum_i k_i^3)$ is enhanced or suppressed in proportion to $\mathcal{O}(1)$ (or larger) powers of $(-k\tau_0)$ on the superhorizon scales, $|u| = (-k\tau_0) \ll 1$, unless $\text{Re}[\mathcal{Q}_0(u)] \sim u^{-1}$ and $\text{Re}[\mathcal{S}_0(u)] \sim u$ which are realized for $\lambda \sim 15$. In particular, by taking the limit $\lambda \to 15$, the above reproduces the result in eq. (3.11):

$$\text{Re}[S_0(|u|)] \simeq \frac{\pi}{2|u|}, \ \text{Re}[Q_0(|u|)] \simeq \frac{\pi|u|}{4}, \ \mathcal{B}_1 = -27, \mathcal{C}_2 = -36x_1^2. \quad \text{(C.26)}$$

Let us comment on how $\lambda \simeq 15$ corresponds to the case of the nearly scale independent $f_{\text{NL}}^{(\text{local,equil})}$. Hereafter, we suppress all indexes about wavenumbers ($\mathbf{k}$) and helicitiy modes ($s_i$), i.e., the three-point function is given by

$$\langle hhh \rangle = i\int_{-\infty}^{\tau_0} d\tau \langle 0|\left[H_{\text{int}}^{\text{PV}}(\tau), hhh\right]|0\rangle, \quad \text{(C.27)}$$

which relates to $f_{\text{NL}}$ as

$$\langle hhh \rangle \propto \delta^{(3)}(\vec{k}_1 + \vec{k}_2 + \vec{k}_3)\frac{\mathcal{A}}{k_1^3 k_2^3 k_3^3}, \ f_{\text{NL}} = \frac{\mathcal{A}}{\sum k_i^3}. \quad \text{(C.28)}$$

For simplicity, let us consider the equilateral triangle and study the interaction Hamiltonian corresponding to $\mathcal{I}_0$ in our paper as

$$H_{\text{int}}^{\text{PV}} \propto -\int \frac{f(\tau)}{a^2}(kh')(kh')(kh'). \quad \text{(C.29)}$$

The generalization to the full interacting Hamiltonian is straightforward. The contribution from the above interaction Hamiltonian to $f_{\text{NL}}$ is

$$f_{\text{NL}} \propto \frac{k_1^4 k_2^4 k_3^4}{\sum k_i^3}\text{Im}\left[h^* h^* h^* \int \frac{f(\tau)d\tau}{a^2} h'h'h'\right], \quad \text{(C.30)}$$

where we ignored real constant prefactors. We arrive at (C.30) by using the definition of $f_{\text{NL}}$ and the action for cubic Weyl term, without assuming any cosmological evolution. Now we evaluate the above in the matter bounce scenario. The derivative of the tensor fluctuation with respect to the conformal time takes a schematic form as

$$h'h'h' = k^{-3/2}\frac{e^{-ik\tau}}{\tau^{12}}\sum_{j=0}^{6}\mathcal{K}_j(k\tau)^j, \quad \text{(C.31)}$$

from which one can derive the following expression of $f_{\rm NL}$,

$$\begin{aligned}
f_{\rm NL} &\propto k^9 {\rm Im}\bigg[k^{-\frac{9}{2}}\int_{-\infty}^{\tau_0} k^{-\frac{9}{2}}\tau^{\lambda-16}d\tau e^{-ik\tau}\sum_{j=0}^{6}\mathcal{K}_j(k\tau)^j\bigg]\\
&\propto k^{15-\lambda}{\rm Im}\bigg[\int_{-\infty}^{-k\tau_0}u^{\lambda-16}du e^{-iu}\sum_{j=0}^{6}\mathcal{K}_j(k\tau)^j\bigg]\\
&= k^{15-\lambda}{\rm Im}\bigg[\sum_{j=0}^{6}\mathcal{K}_j i^{16-\lambda-j}\Gamma(\lambda+j-15,-ik\tau_0)\bigg],
\end{aligned} \qquad (\text{C.32})$$

where we again ignored real constant prefactors. Since we know that $\lambda \simeq 15$ are the special cases, let us inspect the above equation for $\lambda \simeq 15$, particularly for $14 < \lambda < 16$. In this region, $f_{\rm NL}$ has a scale dependence $k^{15-\lambda}$ for $0 < -k\tau_0 \ll 1$, and $f_{\rm NL}$ is scale invariant when $\lambda = 15$, which is a consequence of a direct evaluation of Gamma functions. Away from the region of $14 < \lambda < 16$, the explicit scale dependence depends on each value of $\lambda$. While it is not easy to figure out the scale dependence of $f_{\rm NL}$ from (C.32), the above argument serves as an intuitive way to understand our result. The number $\lambda = 15$ appears in straightforward computation, instead of from any fundamental symmetries.